\documentclass{ws-procs9x6}

\begin{document}

\title{Effects of halo triaxiality, anisotropy and small scale
clumping on WIMP direct detection exclusion limits}

\author{Anne M Green}

\address{Physics Department, 
Stockholm University,\\ 
Stockholm, 10691, SWEDEN\\ 
E-mail: amg@physto.se}


\maketitle

\abstracts{Weakly Interacting Massive Particle (WIMP) direct detection
experiments are closing in on the region of parameter space where
neutralinos may constitute the Galactic halo dark matter. Numerical
simulations and observations of galaxy halos indicate that the
standard Maxwellian halo model is likely to be a poor approximation to
the dark matter distribution. We examine how halo models with
triaxiality and/or velocity anisotropy affect exclusion limits, before
discussing the consequences of the possible survival of small scale
clumps.}

\section{Milky Way halo modelling}

Current Weakly Interacting Massive Particle (WIMP) direct detection
experiments are just reaching the sensitivity required to detect
Galactic dark matter in the form of neutralinos.  The direct detection
event rate and its energy distribution are determined in part by the
WIMP speed distribution, with data analyses nearly always assuming a
standard spherical halo model with isotropic Maxwellian
velocity distribution.

Observational constraints on the form of dark matter halos depend on
the relation of luminous tracer populations to the underlying density
distribution, and are complicated by galactic structure and projection
effects, however it appears that galaxy halos are triaxial (for a
review see Ref.~\cite{sackett}). In particular for the Milky Way,
analysis of local stellar kinematics gives an estimate for the
short-to-long axis ratio of $0.7 \pm 0.1$~\cite{gal1}, while the great
circle tidal streams observed from the Sagittarius dwarf galaxy rule
out a ratio of less than 0.7 in the outer halo~\cite{gal2} (in a
flattened potential angular momentum is not conserved, so that orbits
precess and tidal streams lose their coherence).  Given the
difficulties involved in `observing' dark matter halos it makes sense
to turn to numerical simulations for information on their possible
properties.  In CDM cosmologies structure forms hierarchically; small
objects (often known as subhalos) form first, with larger objects
being formed progressively via mergers and accretion. The shape and
internal structure of galaxy size halos are then determined by the
dynamical processes which act on the component subhalos.

The shape of simulated halos varies, not just between different halos
of the same mass, but also as a function of radius within a single
halo, strongly if the halo has undergone a major merger relatively
recently. Two high resolution Local Group halos studied in detail in
Ref.~\cite{mooredm} have axis ratios of $1:0.78:0.48$ and
$1:0.45:0.38$ at the solar radius. Adding dissipative gas to
simulations tends to preserve the short-to-long axis ratio while
increasing the intermediate-to-long axis ratio~\cite{gas}.

The anisotropy parameter $\beta(r)$, defined as
\begin{equation}
\label{defbeta}
\beta(r)= 1 - \frac{<v_{ \theta}^2>+<v_{ \phi}^2> }{2 <v_{{\rm r}}^2>} \,,
\end{equation}
where $<v_{{\rm \theta}}^2>$, $<v_{{\rm \phi}}^2>$ and $<v_{{\rm
r}}^2>$ are the means of the squares of the velocity components, also
varies with radius.  Typically $\beta(r)$ grows, although not
monotonically, from roughly zero in the centre of the halo to close to
one at the virial radius, with non-negligible variation between
halos. The high resolution galactic mass halos studied in
Ref.~\cite{moore99} have $\beta(R_{\odot})$ in the the range 0.1-0.4,
which corresponds to radially biased orbits.

\section{Effect on exclusion limits}
The differential WIMP event rate due to scalar interactions can be
written in terms of the WIMP scattering cross section on the proton,
$\sigma_{{\rm p}}$:
\begin{equation}
\frac{{\rm d} R}{{\rm d}E} = \zeta \sigma_{{\rm p}} 
              \left[ \frac{\rho_{0.3}}{2}
             \frac{ (m_{{\rm p}}+ m_{\chi})^2}{m_{{\rm p}}^2 m_{\chi}^3}
             A^2 F^2(q)  \int^{\infty}_{v_{{\rm min}}} 
            \frac{f_{v}}{v} \, {\rm d}v 
          \,    \right] \,,
\end{equation}
where the local WIMP density, $\rho_{\chi}$, is normalised to a
fiducial value $\rho_{0.3} =0.3 \, {\rm GeV/ cm^{3}}$ such that
$\zeta=\rho_{\chi} / \rho_{0.3}$, $A$ and $F(q)$ are the atomic number and
form factor of the target nuclei, $f_{v}$ is the WIMP speed
distribution in the rest frame of the detector, normalised to unity,
and $v_{{\rm min}}$ is the minimum WIMP speed which can produce a
recoil of energy $E$: $v_{{\rm min}}=\left( E (m_{\chi}+m_{A})^2 / 2
m_{\chi}^2 m_{A} \right)^{1/2}$.

\begin{figure}
\begin{center}
\includegraphics [width=.495\textwidth]{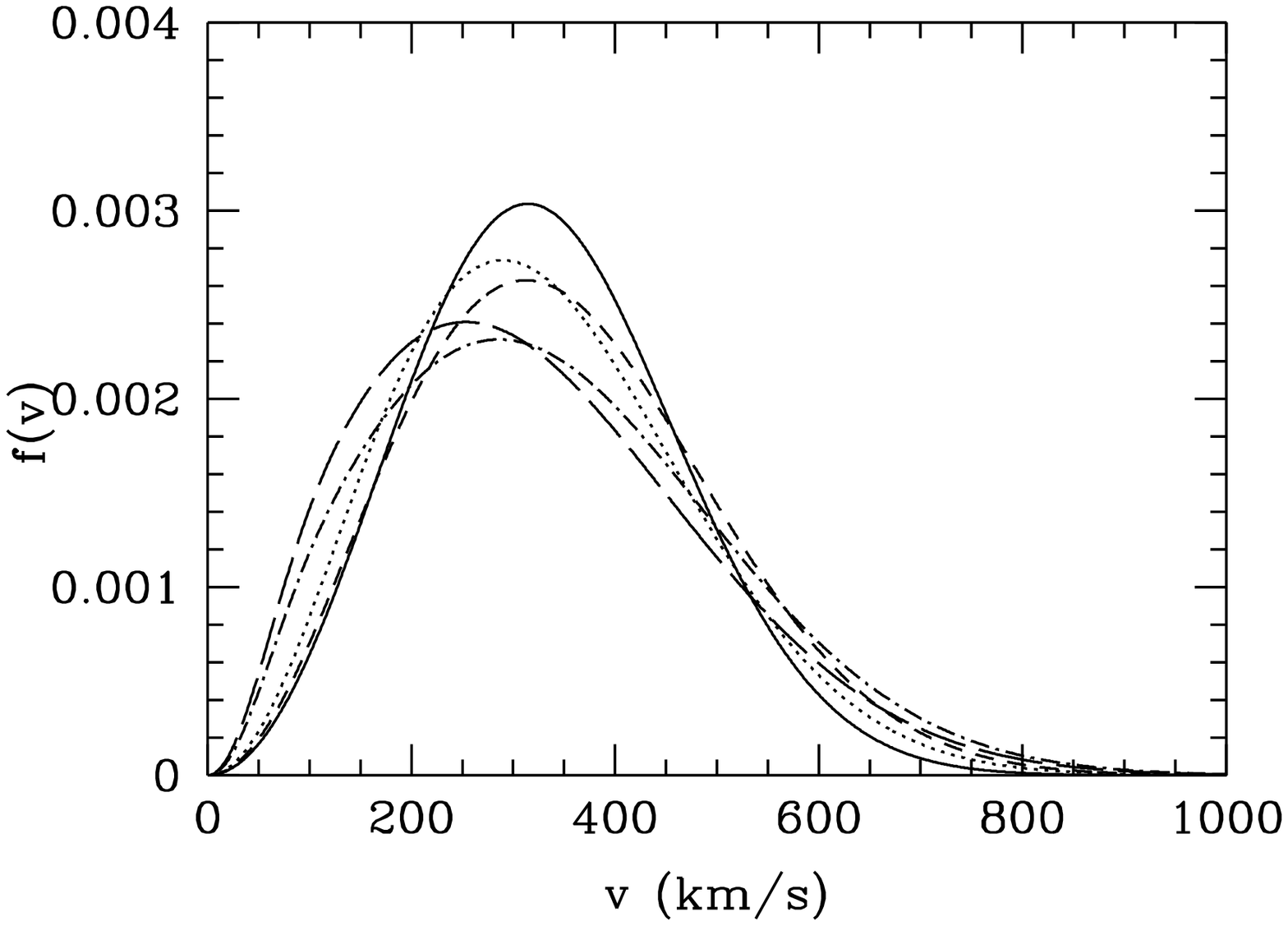}
\includegraphics [width=.495\textwidth]{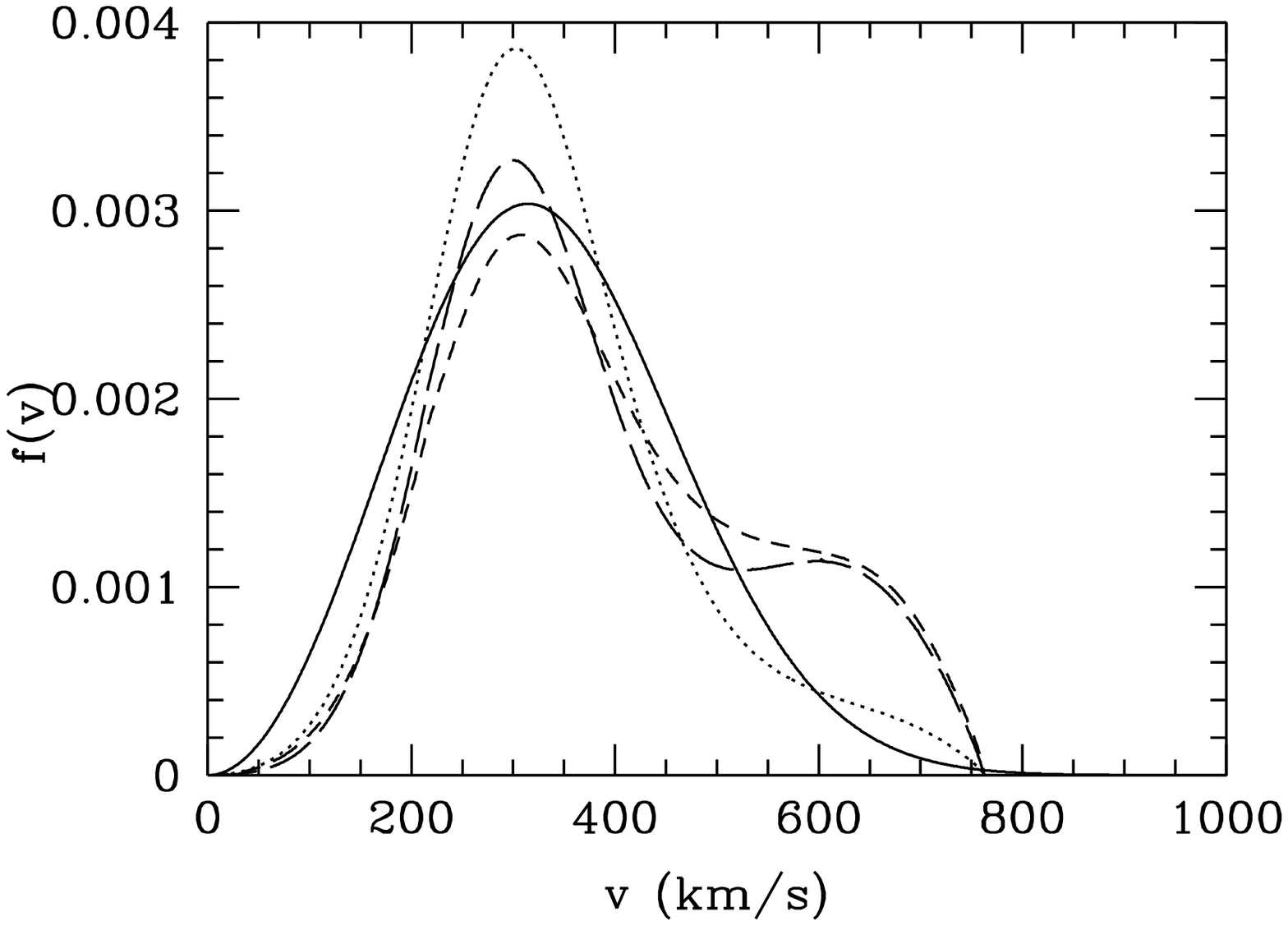}
\end{center}
\caption{The speed distributions, in the rest frame of the Sun, for
the standard halo model (solid line), and the logarithmic ellipsoidal
model on the intermediate axis (left panel) for parameters which
produce axis ratios $1:0.78:0.48$ and $\beta=0.1/0.4$ (dotted/short
dashed) and for axis ratios $1:0.45:0.38$ and $\beta=0.1/0.4$ (long
dashed/dot dashed) and for the OM model (right panel) with
anisotropy radii which produce $\beta(R_{\odot})=0.13, 0.31$ and $0.4$
(dotted, short-dashed, and long-dashed). \label{fig1}}
\end{figure}

We will consider the two self-consistent triaxial and/or anisotropic
halo models which have been studied in relation to WIMP direct
detection to date: the logarithmic ellipsoidal model~\cite{newevans}
and the Osipkov-Merritt anisotropy model~\cite{OM}, studied in
Ref.~\cite{uk}.  The logarithmic ellipsoidal model~\cite{newevans} is
the simplest triaxial generalisation of the isothermal sphere and on
either the long or the intermediate axis the velocity distribution can
be approximated by a multi-variate gaussian. Speed distributions
on the intermediate axis, in the rest frame of the Sun normalised to
unity, are plotted in Fig.~1 along with that for the standard
Maxwellian halo model. The logarithmic ellipsoidal model has a wider
spread in speeds than the standard model, so that the differential
event rate decreases less rapidly with increasing recoil energy.

In the Osipkov-Merritt (OM) model~\cite{OM}, which assumes a
spherically symmetric density profile, the velocity anisotropy varies
as a function of radius as $\beta(r) = r^2 / (r^2 + r_{{\rm a}}^{2})$
so that the degree of anisotropy increases with increasing radius, as
is found in numerical simulations. Following Ref.~\cite{uk} we assume
a NFW density profile with scale radius $r_{{\rm s}}=20$ kpc. We use
values of the anisotropy radius $r_{{\rm a}}$ which correspond to
$\beta(R_{\odot}) = 0.14, \, 0.31$ and $0.4$. The resulting speed
distributions are plotted in Fig.~1.

\begin{figure}
\begin{center}
\includegraphics [width=.495\textwidth]{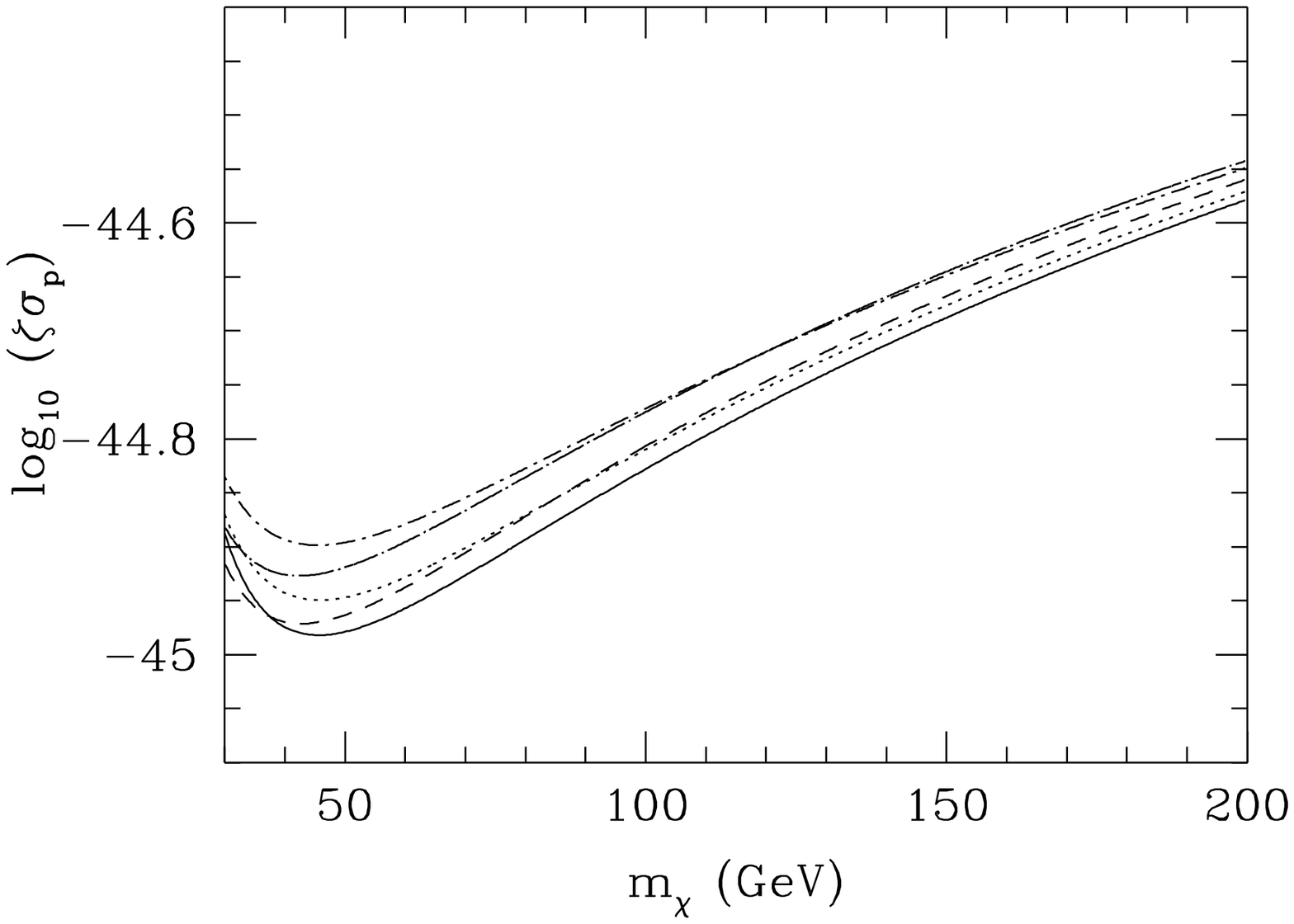}
\includegraphics [width=.495\textwidth]{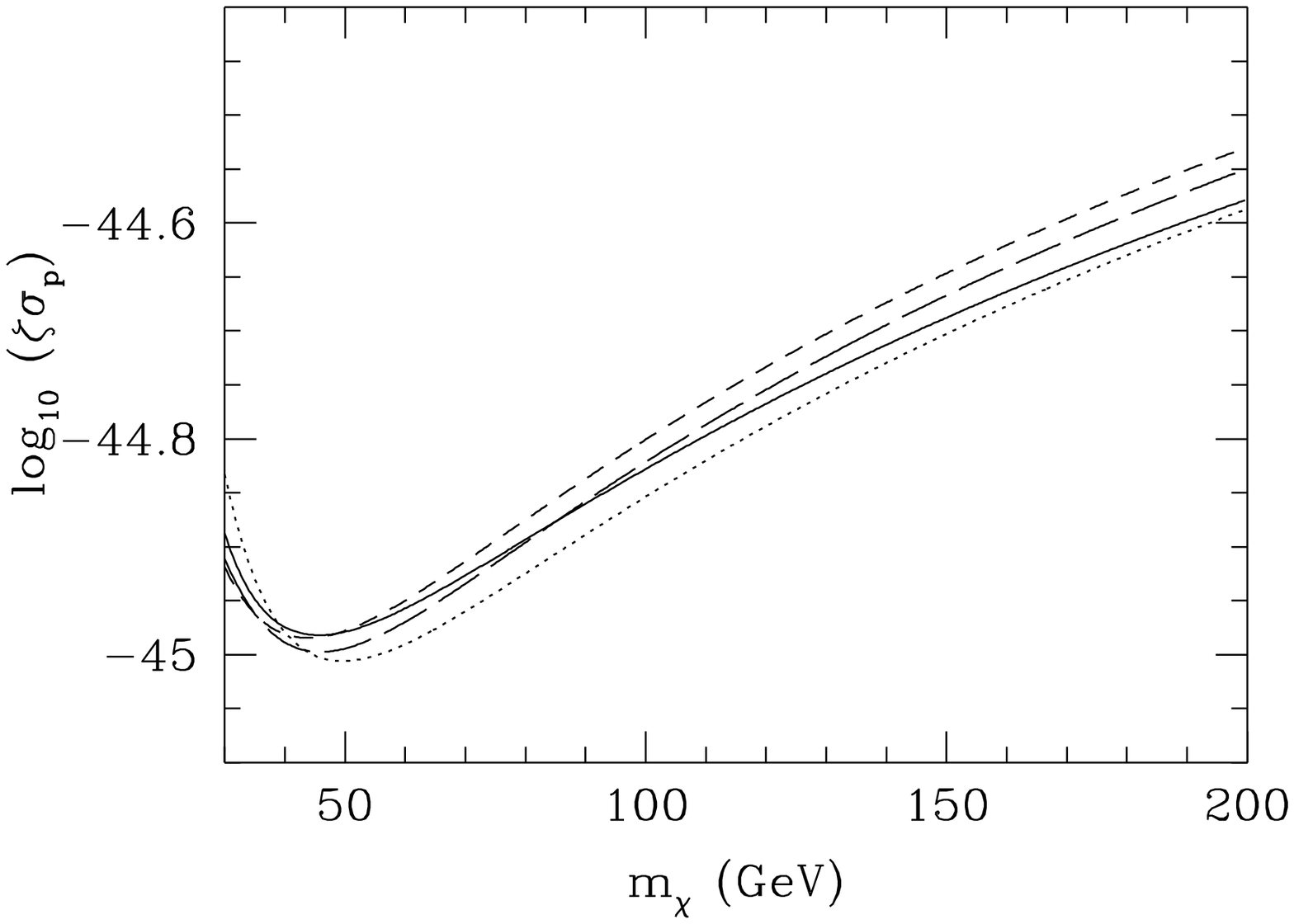}
\end{center}
\caption{The exclusion limits from the IGEX
experiment for the logarithmic ellipsoidal model, location on the
intermediate axis (left panel) and for the OM model
(right panel).}
\end{figure}

In Fig.~2 we plot the exclusion limits found from the IGEX
data~\cite{IGEX}, taking into account the detector resolution, for the
logarithmic ellipsoidal model and for the OM model. We also plot the
exclusion limits from the Heidelberg-Moscow (HM) experiment~\cite{HM}
for the OM model in Fig.~3. Comparing Figs.~2 and 3 we see that the
change in the exclusion limits depends not only on the halo model
under consideration, but also on the data being used; for IGEX the
change in the exclusion limits is largest for large $m_{\chi}$, while
for HM the change is largest for small $m_{\chi}$. This is because for
different $m_{\chi}$, different energy ranges can be most
constraining; for the IGEX data the lowest energy bin is always the
most constraining, while for HM as $m_{\chi}$ increases the constraint
comes from higher energy bins. It should therefore be borne in mind
when comparing exclusion limits from different experiments, that
changing the assumed WIMP speed distribution will affect the limits
from different experiments differently.

The changes in the exclusion limits are not huge (of order tens of
per-cent) for the experiments we have considered, however these
experiments are not optimised for WIMP detection. The change in the
differential event rate, and hence the exclusion limit, would be
larger for an experiment with better energy resolution
and lower threshold energy (see Ref.~\cite{greennew}). We have also
seen that different models with the same value for the anisotropy
parameter $\beta$ can have very different speed distributions, and hence a
different effect on the exclusion limits. Furthermore it is
conceivable that the local WIMP speed distribution may deviate even
further from the standard Maxwellian distribution than the models
that we have considered.

\begin{figure}
\begin{center}
\includegraphics [width=.495\textwidth]{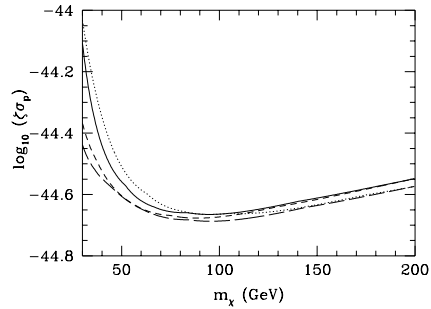}
\end{center}
\caption{The exclusion limits from the HM experiment for the OM
model.}
\end{figure}

\section{Small scale clumping}

Numerical simulations are an extremely powerful tool for understanding
the large scale structure of the universe, however the local dark
matter distribution, which is crucial for direct detection
experiments, can not be probed directly by cosmological
simulations. The smallest subhalos resolvable in the highest
resolution simulations have mass of order $10^{7} M_{\odot}$, while
the first bound neutralino clumps to form would have mass, at
matter-radiation equality, $\sim 10^{-7} M_{\odot}$~\cite{hss}. Other
approaches~\cite{swf,mooredm,hws} have therefore been used to address
the problem, with the results obtained depending on the method used to
extrapolate to small scales below the resolution limit of the
cosmological simulations~\footnote{See Ref.~\cite{greennew} for
a more extensive discussion.}.  Moore et al. claim that the phase space
distribution at the solar radius will depend crucially on the Galaxy's
merger history and on the internal structure of the smallest subhalos,
arguing that it is possible that the local dark matter density could
be zero, or that a single dark matter stream with small velocity
dispersion could dominate, or that many tidal streams could overlap to
give a smooth distribution.

We will now discuss the consequences for the WIMP direct detection
rate if small dense clumps survive at the solar radius. We emphasise
that it is not clear at the moment whether this is the case, and
further work is necessary to clarify this issue. We could then be
currently located within a clump with local density in excess of the
mean value, on the other hand it is possible that we could be in a
region between clumps where the WIMP density is
zero~\cite{mooredm}. In the latter case all current attempts at WIMP
direct detection would be doomed to failure~\footnote{WIMPs could
still be detected using ancient mica though as it has an integtation
time of order a Gyr~\cite{mica}.}, and exclusion limits would tell us
nothing about the WIMP cross-section. At the other extreme a dense
clump at the earth's location would produce an enhanced event rate and
a distinctive experimental signal (the theoretical differential event
rate would be a step function with amplitude inversely proportional to
the speed of the subhalo with respect to the earth, and position
increasing with increasing relative speed and WIMP mass). For small
clump velocities and WIMP masses there would be no constraint on the
WIMP cross-section (no WIMPs would have sufficient energy to cause an
observable recoil), but as the WIMP mass is increased the constraints
would become much tighter as then all the WIMPs would be energetic
enough to cause events of a given recoil energy.

\section{Conclusions}
We have seen that even if the local WIMP distribution is smooth its
velocity distribution may deviate significantly from the standard
Maxwellian, and this has a non-negligible effect on exclusion limits
from WIMP direct detection experiments, affecting the limits from
different experiments differently. The effect on experiments which can
detect the angular and/or time variation of the event rate will be
more significant. Constraints (and in the future possibly best fits)
calculated assuming a standard Maxwellian halo could be erroneous,
even worse if only the signals expected from the standard halo model
(e.g. a sinusoidal annual modulation with peak in June) are searched
for, a real WIMP signal could be overlooked. On the other hand, more
optimistically, if WIMPs were detected it might then be possible to
derive useful information about the local velocity distribution, and
hence the formation of the Galactic halo.

\end{document}